\documentclass[twocolumn,showpacs,preprintnumbers,amsmath,amssymb]{revtex4}

\usepackage{graphicx}
\usepackage{dcolumn}
\usepackage{bm}
\usepackage{here}

\begin{document}

\preprint{Revtex4}

\title{Controllability of Wavepacket Dynamics in Coherently Driven Double-Well Potential
}

 \author{Akira Igarashi}
 \email{f99j806b@mail.cc.niigata-u.ac.jp}
 \affiliation{
Graduate School of Science and Technology, 
 Niigata University, Ikarashi 2-Nochou 8050, Niigata 950-2181, Japan
}

\author{Hiroaki Yamada}
 \email{hyamada@uranus.dti.ne.jp}
\affiliation{%
YPRL,  5-7-14 Aoyama, Niigata 950-2002, Japan
}%
\date{\today}

\begin{abstract}
 We numerically study  controllability of quantum dynamics 
  in a  perturbed one-dimensional double-well potential by using an optimal control theory.
  When the perturbation strength is small the dynamics of  an initially localized Gaussian wavepacket 
shows  coherent oscillation between the wells.
 It is found that  as the increase of the strength and/or the number of frequency components of
  the perturbation the coherent motion of the Gaussian wavepacket 
changes to an irregular one with irreversible delocalization.
We  investigate the controllability of the system depending on the perturbation parameters 
and the initial quantum state by focusing mainly on the delocalized state generated 
by the polychromatical perturbation.
In  relatively long-time control for the Gaussian wavepacket and the delocalized state,
we show that it 
is well-controllable via the first excited state doublet in spite of the perturbation parameters.
 On the other hand,  in the relatively short-time control we show  difficulty of the control 
for the delocalized state because of the numerous local minima.
Furthermore, it is demonstrated that chaotic phase space structure of the system assists the
 controllability of the delocalized state.
\end{abstract}


\maketitle

\section{Introduction}
Decoherence of the quantum 
dynamics and  a  control  of the wavepacket 
dynamics by using  an  external force field are recent attractive topics.
There are many studies on environment-induced quantum
   decoherence by coupling the quantum system to a reservoir
   \cite{leggett81,weiss99,dittrich86,choen99}.
Quantum dissipation due to the interaction with chaotic degrees
   of freedom has been also studied \cite{zurek81,kolovsky94,percival98}.
It is expected that the quantum dynamics is sensitive to any disruption of coherence as it occurs due to 
 an  unavoidable coupling to the environment.
One of obstacles for  experimental realization of a quantum computer
is decoherence of a quantum state caused by coupling to the other degrees of freedom \cite{nielsen00}.
 
On the other hand, in the experimental realization of control of the chemical reactions and 
the optical molecular devices,  
a design of the control field based on optimal control theory (OCT),
is also an important topic \cite{rice00,shapiro03,zhu98,umeda01,ohta00}.
  Recently, some methods to control the transition between  quantum states 
have been proposed \cite{takami05,dey00,dannis03,bbrgmann98,zhu01,schrimer01,dearaujo03,sugawara03,alessandro02}.
Our main interest is in 
the controllability between the quantum states in, what we call, a quantum chaos system 
\cite{gutzwiller90,haake01}.

In the previous papers \cite{igarashi05,igarashi06a}, 
we investigated influence of chaos on quantum tunneling and decoherence in a parametrically
driven double-well system modeled by  the  following Hamiltonian, 
\begin{eqnarray}
      H_0(t) & = & \frac{p^2}{2} + \frac{q^4}{4} -A(t) \frac{q^2}{2},  \\
      A(t) & =& a-\frac{\epsilon}{\surd M} \sum_{i=1}^M \sin(\Omega_i  t ),   
\end{eqnarray}
where $q$ and $p$ represent the generalized coordinate and the conjugate momentum, respectively.
Here $M$ denotes the number of  mutually incommensurate frequency components
 $\{ \Omega_i, i=1,2,...,M \}$. 
We choose off-resonant frequencies which are 
   far from both classical and quantum resonance
   in the corresponding unperturbed case ($\epsilon=0$).
The number $M$ of  frequency components in the polychromatic oscillation 
corresponds to the number of degrees of freedom coupled to the unperturbed  double-well system
\cite{frequency}.
 The details of relation between the time-dependent Hamiltonian and the autonomous representation
have been shown in Ref.\cite{yamada99}.

Note that in the previous works of Lin {\it et al.}, they 
dealt with a double-well system driven by single forced oscillator, 
therefore, the asymmetry of the potential play a role in the chaotic behavior and tunneling transition between the
quasienergy states \cite{lin90,grossmann91,gammaitoni98}. 
However, in our model the potential form is remained symmetric during the time evolution process,  
 and different mechanism from the forced oscillation makes 
the classical chaotic behavior. 
We classified the motions of an  initially localized Gaussian wavepacket depending on the 
perturbation parameters ($M, \epsilon$), i.e.  {\it coherent motions} like an instanton and {\it irregular motions}
of a delocalized state \cite{igarashi06a}.  
We call the delocalized state in the irregular motion {\it chaos-induced delocalized state} in this paper.

The purpose of this article is to present  numerical results 
 on  the control of the quantum dynamics
in the one-dimensional double-well potential with the coherent perturbation.
The controllability of the system depends on the perturbation parameters and the initial state.
In the relatively long-time control the displacement of the Gaussian wavepacket 
is well-controllable by a designed field based on OCT 
  in the perturbed double-well potential 
regardless of the external parameters.
  Furthermore, in the relatively short-time control  we show 
the difficulty of the control for the chaos-induced delocalized state 
generated under the perturbation because of numerous local optimal fields, 
compared to the case of Gaussian wavepacket.
It is numerically demonstrated that 
when the system is fully chaotic by the increase of the perturbation strength $\epsilon$ 
the controllability of the delocalized state can be enhanced by the chaotic behavior. 
We call the phenomenon {\it chaos-assisted control}  in this paper.  

 The remained sections of the present paper are as follows.
In the next section we give some backgrounds and the numerical setting of the model
and  brief review of  chaos-induced delocalized state. 
In Sect. 3 we shortly explain the optimal control theory used in the numerical calculation.
In Sect. 4 we give  numerical results for control between Gaussian wavepackets
in the polychromatically perturbed system.
In Sect. 5 we show  results of the transition between  a chaos-induced delocalized state 
and  a Gaussian wavepacket. 
In Sect. 6 and 7 the difficulty of the control  due to  the local minima and the chaos-assisted control 
are investigated for the short-time control, respectively.
The last section contains summary and discussion.

\section{Model and chaos-induced delocalized state}
In this section we give the details of the setting in the Eq.\~(2) and in  the numerical calculation.
Furthermore,  we give a brief review on  occurrence of 
 a  irregular motion in the delocalized state.
The  parameters are set as $a=5$, $\epsilon=0.1 \sim 1.0$, 
to emphasize the tunneling effect in the energetically and/or dynamically forbidden region
during the time evolution.
We used  the following  Gaussian wavepacket localized at 
a bottom of the right well  as, 
\begin{eqnarray}
  \psi_i (q) =   (\sigma \pi)^{1/4}  \exp \{ -\frac{(q-q_0)^2}{2\sigma}  \}, 
\end{eqnarray}
where  $q_0=\surd a$ is the postion of the right bottom 
and  the  spread of the initial packet $\sigma \sim 0.3$  as the initial state at $t=0$.  
 $\hbar=1.0$ and time mesh $\delta t$ is order of $10^{-2}$. 
The spatial discretization is chosen so that $2^8(=256)$ points cover
the interval  $[-5.5,5.5]$.  
 The Gaussian wavepacket can be approximately generated by the linear combination
of the ground state doublet as $ \psi_i (q) \sim \frac{1}{\surd 2}(\phi_0(q) + \phi_1(q)  )$, where
$\phi_0$ and $\phi_1$ denote the ground state doublet. (See Fig.1.)
  Indeed, the ammonia molecule is well described by this barrier height
  with two doublets
with energies below the top of the barrier, 
and the quantum dynamics is well approximated by 
 the several levels 
from the bottom of the well in the weekly perturbed case \cite{delgado02}.

   \begin{figure}[!ht]
    \centering
  \includegraphics[clip,scale=1.2]{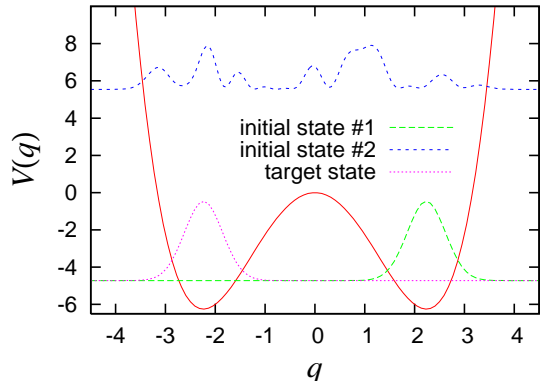}
    \caption{ 
The initial (target) Gaussian wave function $|\psi_i(t=0)|^2$ ($|\psi_f(t=t_f)|^2 \equiv |\Phi_f|^2$ ), respectively. 
The dotted line shows a  delocalized state as  the  initial state, which is generated by 
time-evolution ($t=5.54 \times 10^3$) in the perturbed case ($M=10,\epsilon=1.0$).
The curve of the double-well potential is also plotted.
    }
   \end{figure}

In the classical dynamics, such a system shows chaotic behavior by the oscillatory 
force $A(t)$ \cite{igarashi05,igarashi06a}.
The Newton's equation of the motion in the monochromatically perturbed case ($M=1$) is 
\begin{eqnarray}
  \frac{d^2 q}{dt^2} -(a-\epsilon \sin \Omega t)q +q^3=0.
\end{eqnarray}
\noindent
Note that the classical system is known as nonlinear Mathieu equation
which can be derived from surface acoustic wave in piezoelectric solid \cite{konno90} 
and nanomechanical amplifier in micronscale devices \cite{harrington02}.
In the polychromatically perturbed cases ($M>1$) the smaller the strength $\epsilon$ can generate 
chaotic behavior of the classical trajectories the larger $M$ is \cite{igarashi06a}.



   We define transition probability $P_L(t)$ of finding the wave packet in the left well and 
a {\it degree of coherent motion}, 
   $\Delta P_{L}$, based on the fluctuation of the  transition probability as,  
   \begin{eqnarray}
    P_{L}(t) & \equiv &   \int_{-\infty}^{0}|\psi(q,t)|^{2}\,dq,  \\
    \Delta P_{L}  & \equiv & \surd \langle (P_{L}(t)-\langle P_L(t) \rangle_{T})^{2} \rangle _T, 
   \end{eqnarray}
   where $\langle ... \rangle _{T}$ represents time average value for a period 
   $T (\sim 9.4 \times 10^3)$ \cite{igarashi06a}.
   In the cases that perturbation strength $\epsilon$ is relatively small, 
   $P_L(t)$ can be interpreted as the tunneling probability that the initially
   localized wave packet goes through the central energy barrier and reaches
   the left well.

   \begin{figure}[!ht]
    \centering
\includegraphics[clip,scale=0.67]{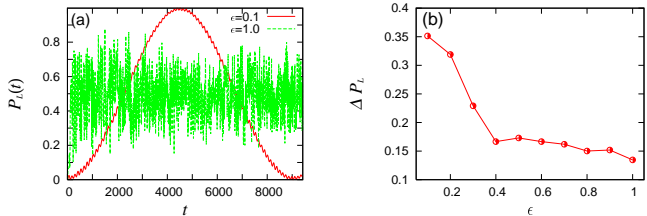}
    \caption{
(a)Transition probability $P_{L}(t)$ as a function of time $t$
    for $M=10$ with $\epsilon=0.1,1.0$. 
(b)$\epsilon$ dependence of the degree of the coherent motion  $\Delta P_L$ in the case.
    }
   \end{figure}

   Figure 2(a) shows the time-dependence of 
   $P_{L}(t)$ in the polychromatically perturbed case ($M=10$).
   Apparently we can observe two kinds of  time-dependence, 
i.e.  coherent ($\epsilon=0.1$) and irregular ($\epsilon=1.0$)  motions.
In order to estimate quantitatively the difference 
   between coherent and  irregular  motions, 
 we use the  {\it degree of coherent motion}  $\Delta P_L$ in Eq.\~(6) and 
roughly divide the types of  motion of wavepacket as seen in Fig.\~2(b).
   In the {\it coherent motions}, the values  of $\Delta P_{L}$'s are
   almost same to the unperturbed case, i.e. $\Delta P_{L} \gtrsim 0.3$. 
   In the {\it irregular motions} which are similar to the stochastically perturbed case, 
   the values  of $\Delta P_{L}$'s becomes much smaller, i.e. 
   $\Delta P_{L} \lesssim 0.2$. 
In the irregular motions, the wavepacket is delocalized over the
both the wells as seen in Fig.1. 
The delocalized state
cannot show the recurrence  to the Gaussian shape within the accessible computation time. 
We call  delocalized states in the irregular motions 
{\it chaos-induced delocalized state} in this paper.


  
It is well-known that  an existence of  a turning point breaks 
the Gaussian shape during the tunneling process even in  a  coherent motion.
However, in the coherent motion after the tunneling process the wavepacket goes back to the Gaussian shape again. 
In the following sections we give numerical results  of 
the controllability for the  Gaussian wavepacket and 
the {\it chaos-induced delocalized state}
corresponding to the irregular motion.

\section{Optimal Control of Quantum Dynamics}
In this section, we briefly explain the OCT \cite{rice00,zhu98} we used in the numerical calculation. 
 
We consider  the  total Hamiltonian $H(t)$ combined the system $H_0(t)$ with
the interaction by the external field $E(t)$ as, 
\begin{eqnarray}
      H (t) & = & H_0(t)-\mu(q)E(t).
\end{eqnarray}
The second term in Eq.\~(7) means  the  interaction between the transition dipole moment
 $\mu(q)( \equiv q)$ and  the external field $E(t)$, where the charge is set unity. 
 Note that we adapted  the Hamiltonian $H_0(t)$, including the time-dependent part $-A(t)\frac{q^2}{2}$, as
a  controlled system. However, we confirmed that if we use the static double-well system, i.e. $\epsilon=0$, 
as the controlled system the numerical result does not change in the essential point,  except for 
 one in Sect.VII where the effect of the perturbation strength is discussed.

 The OCT gives an optimized external field $E(t)$ in order to produce
  the target state $\Phi_f$ at the target time $t=t_f$
starting from an initial state $\psi_i(t=0)$. 
The objective functional $J[E(t)]$ so as to maximize is 
\begin{eqnarray}
 & &J[E(t)]  =  | \langle \psi_i(t_f)|\Phi_f \rangle |^2 -\alpha \int_0^{t_f} E(t)^2 dt  \nonumber \\
&-& 2Re \Biggl[ \langle  \psi_i(t_f)|\Phi_f \rangle  \int_0^{t_f}  \langle \psi_f(t)|\frac{\partial}{\partial t}-iH(t)|\psi_i(t) \rangle  dt \Biggl],   
\end{eqnarray}
Here  $\psi_i(t)$ is a wave function driven by the optimal field $E(t)$ 
which has been determined by the quantum states, $\psi_i(t)$ and  $\psi_f(t)$ as, 
\begin{eqnarray}
E(t)= \frac{ Im [ \langle \psi_i(t)|\psi_f(t) \rangle \langle \psi_f(t)|\mu(q)|\psi_i(t) \rangle ] }{\alpha} .
\end{eqnarray}
$\psi_f(t)$ is  an inversely evolving quantum state
starting from the target state $\psi_f(t=t_f)=\Phi_f$, 
  which can be interpreted as a Lagrange multiplier to satisfy Schr\"{o}dinger equation.  
The second term in Eq.(8) is put to limit the resulting field intensity, where 
$\alpha$ is a penalty factor to weight the significance of the external field. 
  The optimal field strongly  
depends on the initial and final states, 
the potential parameters and the target time $t_f$.
Note that the existence of the optimal field has been proven 
for large target time and small $\alpha$ limit in the 
unperturbed case \cite{rice00}. 
Generally, when we can optimize the transition between the energy eigenstates 
 by a $\pi-$pulse,  
 the target time must be at least $t_f  \sim O(\hbar/\Delta E)$ to resolve the energy difference
$\Delta E$.

The relation between the controllability and chaotic behavior in  quantum systems 
 has been investigated in a context of OCT \cite{takami05}.

\section{Controllability of the Gaussian Wavepacket}
In this section,  we give  numerical results  that displacement of 
the Gaussian wavepacket can be well controlled 
 in the perturbed double-well potential.

\subsection{long-time control ($t_f=300$)}
We would like to control the transition between  non-stationary states.
 Here we take Gaussian wavepackets localized around $q=q_0$($q=-q_0$) as the 
initial(target) state, respectively. (See Fig.1.)
Accordingly, in the quantum control there is no energy transfer 
between the initial and target states. 
We take the target time $t_f=300$ that is much smaller than a tunneling time scale 
$T\equiv \hbar/\Delta E_{12} \sim 9.7 \times 10^{4}$ 
in the unperturbed case ($\epsilon=0$).

\begin{figure}[h!]
\centering
\includegraphics[clip,scale=0.67]{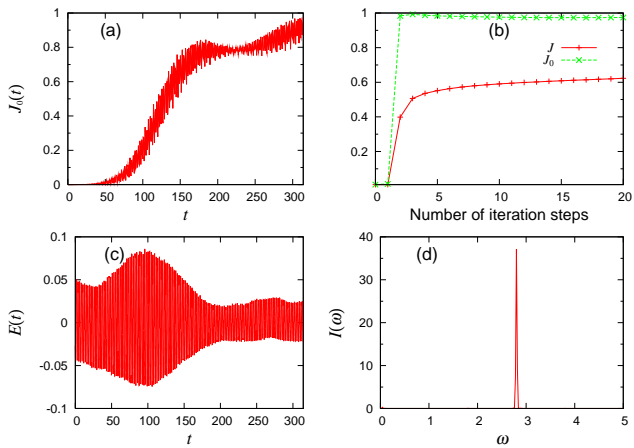}
\caption{
Long-time optimal control of the Gaussian wavepacket
 in the monochromatically perturbed case($M=1,\epsilon=0.1$). 
(a)Time-dependence of the squared final overlap $J_0(t)$ after 20 iterations. 
(b)Convergence of the optimized objective functional values  $J$ and $J_0$ as a function of the
number of iteration.
The panels (c) and (d) show the optimized fields $E(t)$ and the power spectrum $I(\omega)$ 
of the field.
We used $\alpha=1.0$  and $t_f=300$ in the calculation.
}
\label{fig2} 
\end{figure} 

Figure 3(a) shows the final overlap $J_0(t) \equiv | \langle \psi_i(t)|\Phi_f \rangle |^2 $ 
after 20 iterations in the monochromatically perturbed  case ($M=1$) for
 the target time $t_f=300$.
 As we can expect,  based on the OCT, the iteration of the $J_0$ shows fast convergence~\cite{zho98}.
The final state by the optimal field is well overlapped for the target Gaussian function.
  In Fig.3(c) and (d),  time dependence of the optimized field and the power spectrum are shown.
The main peak of the power spectrum corresponds to energy difference between 
the ground state and first excited state doublets  that energy
is lower than the barrier height. 
We can immediately confirm that the wavepacket travels between the wells via the first excited state
doublet by calculating some quantum mechanical mean values,  $<H>, <q>, <p>$, and so on 
during the time-evolution process. 

\begin{figure}[h!]
\centering
\includegraphics[clip,scale=0.67]{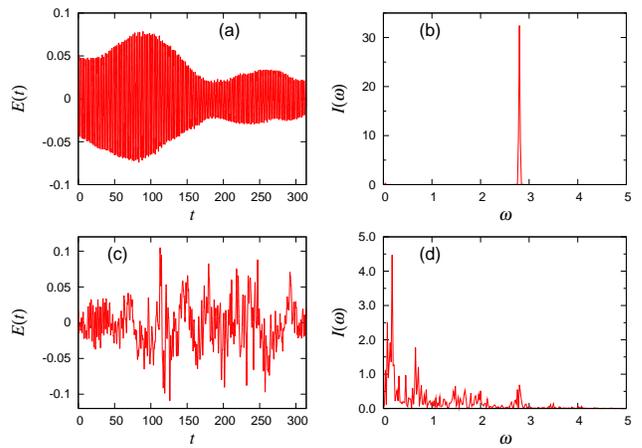}
\caption{
Long-time optimal control of the Gaussian wavepacket
 in the polychromatically perturbed case ($M=10$). 
(a) and (c) are the optimal fields after 20 iterations 
for $\epsilon=0.1$ and $\epsilon=1.0$, respectively.
(b) and (d) are the power spectra for $\epsilon=0.1$ and $\epsilon=1.0$, respectively.
We used $\alpha=1.0$  and $t_f=300$ in the calculation.
$J_0(t_f) \sim 1.0$ in the both cases.
}
\label{fig4} 
\end{figure} 

Figure 4 shows the relatively long-time optimal control for the displacement of the Gaussian wavepacket
 in the polychromatically perturbed cases ($M=10$).
The function form of the optimal field depends on the extent of the chaotic behavior 
in the classical system. 
In the strongly chaotic case ($\epsilon=1.0$) where the perturbation strength is large,  the
optimized field becomes a complicated function of time with many modes.
The peak structure of the power spectrum shows that a lot of modes are necessary to 
shift the Gaussian wave packet in the strongly chaotic cases.

As a result, for the relatively long-time control 
between the Gaussian wavepackets  with no energy deference
it is almost perfectly controlled ($J_0(t_f) \sim 1.0$) regardless of the perturbation type.

\subsection{short-time control ($t_f=30$)}
In this subsection we also consider the control problem between the Gaussian wavepackets in 
the relatively short-target-time $t_f=30$.
It can be expected that the optimal field has much different function form from 
the $\pi-$pulse type for the
short target time $t_f$,  and that 
the local minima may cause  instability of the convergence to the global optimal field.

\begin{figure}[h!]
\centering
\includegraphics[clip,scale=0.67]{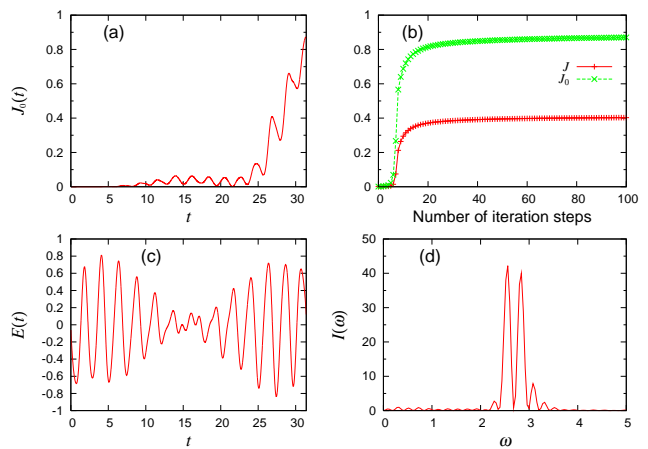}
\caption{
Short-time optimal control of the Gaussian wavepacket 
in the monochromatically perturbed case ($M=1,\epsilon=0.1$). 
(a)Time-dependence of the squared final overlap $J_0(t)$ after 100 iterations. 
(b)Convergence of the optimized objective functional values  $J$ and $J_0$ as a function of the
number of iteration.
The panels (c) and (d) show the optimized field $E(t)$ and the power spectrum $I(\omega)$ 
of the field, respectively.
We used $\alpha=0.1$  and $t_f=30$ in the calculations.
$J_0(t_f) \sim 0.9$ after 100 iterations.
}
\label{fig5} 
\end{figure} 

\begin{figure}[h!]
\centering
\includegraphics[clip,scale=0.67]{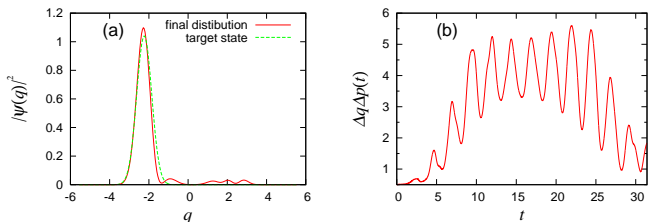}
\caption{
(a) The probability density of the target Gaussian wavepacket and 
the final wavepacket achieved  by  using the optimal field in Fig.5(c).
(b) The time-dependence of the uncertainty product $\Delta p \Delta q$ 
during the time-evolution by the optimal field.
}
\label{fig6} 
\end{figure} 

In Fig.5 we show the result of the short-time optimal control 
in the monochromatically perturbed case ($M=1,\epsilon=0.1$). 
By adjusting the penalty parameter $\alpha$ for the optimization 
the optimal field can be almost found out even for the short-time control ($J_0(t_f) \sim 0.9$).

However, as seen in Fig.6(a) the final wavepacket has uncontrollable tail even after many iterations.
The short-time optimal control is much harder than the relatively long-time ones.
Figure 6(b) shows the time-dependence of the uncertainty product $\Delta p \Delta q 
\equiv  \surd \langle (q-\langle q \rangle )^2 \rangle
     \surd \langle (p-\langle p \rangle )^2\rangle$,   
during the time evolution by the optimal field in Fig.5(c), 
   where $\langle...\rangle $ denotes quantum mechanical average.
We use uncertainty product as a simple measure for quantum fluctuation in the phase space. 
It is found that the Gaussian wavepacket does not delocalize during the process.
Furthermore, we have confirmed that the similar optimal short-time control can be achieved for the other cases
with different perturbation parameters ($M=10,\epsilon=0.1$), ($M=10,\epsilon=1.0$), and so on.
Here, we used $\alpha=0.1$ to adjust the strength of the external field $E(t)$. 
The small value of the penalty factor allows the optimized filed with the large amplitude.
The existence of a lot of  different optimized fields depending on 
$\alpha$  means the existence of a lot of local minima in the solution space.
The details of the $\alpha-$dependence will be given elsewhere \cite{igarashi06c}.

Generally speaking, 
the non-stationary state $\Psi(t)$ at time $t$ can be expressed by a set of the 
expansion coefficients $\{ c_j \}_t$ as $\Psi(t)=\sum_j c_j(t) \phi_j$, where $\{ \phi_j \}$ are eigen states
in the unperturbed case.
Accordingly, the quantum control from the initial state at $t=0$ to the final state
at $t=t_f$ is a problem to answer a question that how we can get the 
 set of the coefficients $\{ c_j \}_{t=t_f}$  in $\Phi_f =\sum_j c_j(t=t_f) \phi_j $ in starting from
$\{ c_j \}_{t=0}$ with obeying the Schr\"{o}dinger equation.
We can see that the transition between the Gaussian wavepackets can be almost attained
by a change of the relatively  {\it simple phase-relation} between the states via the 
first excited state doublet 
as $(c_0(0)=1/\surd 2, c_1(0)=1/\surd 2) \to (c_0(t_f)=1/\surd 2, c_1(t_f)=-1/\surd 2) $.

\section{Controllability of the Delocalized State}
In this section, we investigate the controllability 
of the {\it chaos-induced delocalized state}
 as the initial state $\psi_i(t=0) $ of the optimization procedure.
Note that the target state is a localized Gaussian wavepacket in the left-well
as well as the last section.
Hereafter, we mainly use the polychromatically perturbed case ($M=10,\epsilon=1.0$) 
as a typical controlled system $H_0(t)$, which is strongly chaotic as
shown in Fig.2.  


\begin{figure}[h!]
\centering
\includegraphics[clip,scale=0.67]{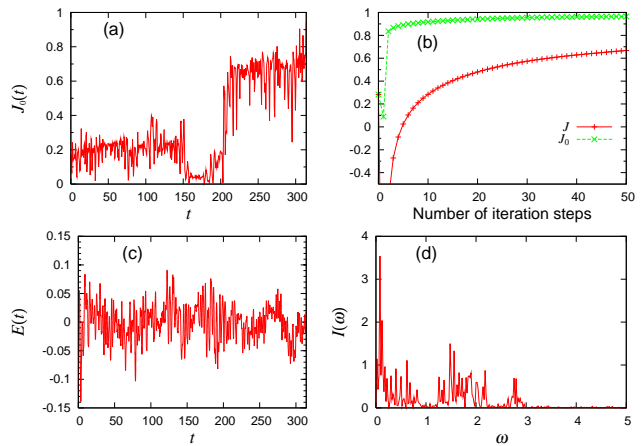}
\caption{
Long-time optimal control for the delocalized initial state  
in the polychromatically perturbed case ($M=10,\epsilon=1.0$). 
(a)Time-dependence of the squared final overlap $J_0(t)$ after 50 iterations.  
(b)Convergence of the optimized objective functional values  $J$ and $J_0$ as a function of the
number of iteration.
The panels (c) and (d) show the optimized fields $E(t)$ and the power spectrum $I(\omega)$ 
of the field.
We used $\alpha=1.0$  and $t_f=300$ in the calculation. 
As an initial state we used the initial delocalized state 
obtained by time-evolution up to $t=5.54 \times 10^3$ in the system with $M=10,\epsilon=1.0$.
}
\label{fig7} 
\end{figure} 

\begin{figure}[h!]
\centering
\includegraphics[clip,scale=0.67]{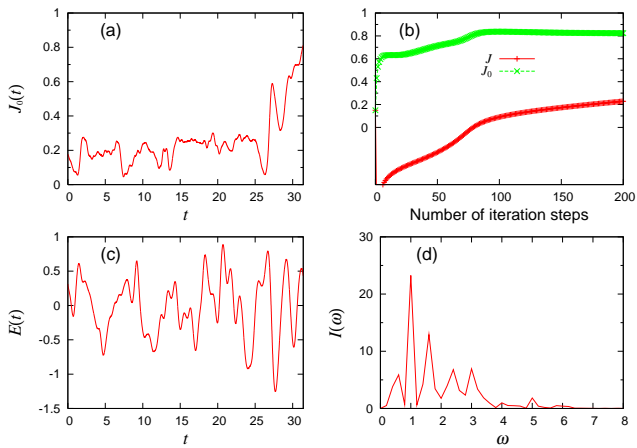}
\caption{
Short-time optimal control for the delocalized initial state  
in the polychromatically perturbed case ($M=10,\epsilon=1.0$). 
(a)Time-dependence of the squared final overlap $J_0(t)$ after 200 iterations.  
(b)Convergence of the optimized objective functional values  $J$ and $J_0$ as a function of the
number of iteration.
The panels (c) and (d) show the optimized field $E(t)$ and the power spectrum $I(\omega)$ 
of the field.
We used $\alpha=0.1$  and $t_f=30$ in the calculation. 
As an initial state we used the initial delocalized state 
obtained  time-evolution up to $t=5.54 \times 10^3$ in the system with $M=10,\epsilon=1.0$.
}
\label{fig8} 
\end{figure} 

Figure 7 shows the relatively long-time control of the delocalized state in 
 the polychromatically perturbed case ($M=10, \epsilon=1.0$). 
The achievement of the optimization is almost same as the case of the Gaussian wavepacket.
Figure 8 shows the short-time control for the same case.
It is difficult to design the  optimal field  for the short target-time ($t_f=30$).
As seen in Fig.8(b), it seems that the objective functional value 
saturates around $J_0 \sim 0.8$ even for 200 iterations, and  the state is trapped in a 
local minimum in the optimization process. 
The optimized field consists of a lot of complicated modes as seen in Fig.8(d).

\begin{figure}[h!]
\centering
\includegraphics[clip,scale=0.67]{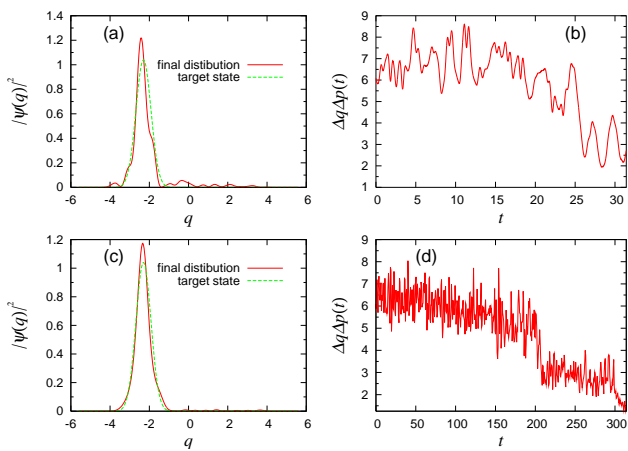}
\caption{
(a), (c) The probability density of the target wavepacket and 
final wavepacket achieved  by  using the optimal field in Fig.8(c) and Fig.7(c), respectively.
(b), (d) The time-dependence of the uncertainty product $\Delta p \Delta q$ 
during the time-evolution by the optimal field in Fig.8(c) and Fig.7(c), respectively.
}
\label{fig9} 
\end{figure} 

\begin{figure}[h!]
\centering
\includegraphics[clip,scale=1.2]{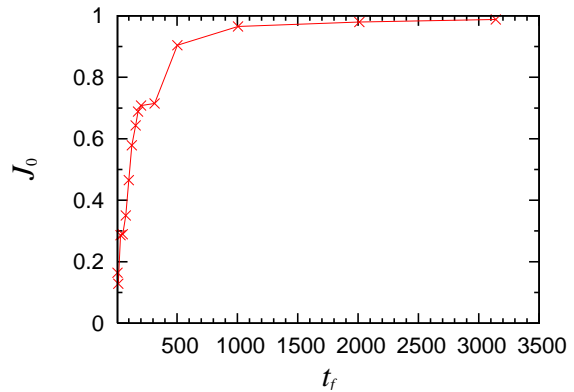}
\caption{
 $t_f-$dependence of the 
optimal control of the delocalized state  
in the unperturbed case.
 The squared final overlap $J_0(t_f)$ after 50 iterations as a function of target time $t_f$.
We used $\alpha=1.0$ in the all calculations.
As an initial state we used the initial delocalized state 
obtained  by  time-evolution up to $t=5.54 \times 10^3$ in the system with $M=10,\epsilon=1.0$.
}
\label{fig10} 
\end{figure} 

Figure 9(a) and (c) show the final states by using the optimized fields in Fig.8(c) and Fig.7(c), respectively.
Figure 9(b) and (d) show the time-dependence of $\Delta p  \Delta q$ during the time-evolution by the
optimized field. 
Although the quantum fluctuation gradually decreases  from the
initial large value 
of the delocalized state
toward small value ($\sim 0.5$) in the final Gaussian wavepacket, 
it is hard to squeeze the delocalized state in the short-time control.

Figure 10 shows the final overlap $J_0$ as a function of target time $t_f$.
 The time-dependence of the functional values $J_0$ show irregular increase with time.
It seems that the $J_0$ suddenly increases around $t_f=100$ and  almost saturates to $J_0 \sim 1.0$
 around $t_f \sim 1000$.
Apparently the longer the target time becomes the better optimization achieves.
We can expect that the optimized field converges to the global minimum
 for the longer target time.

It should be emphasized that even for the long-time control the difference 
of the controllability between the delocalized state and the Gaussian wavepacket 
is the number of iterations 
for the convergence that represents a cost 
for the optimization. The chaos-induced delocalized state consists of 
a lot of eigenstates on the unperturbed basis
due to the mixing of multilevel  caused by the perturbation.

\section{local optimal field}
So far,  we used a typical  sinusoidal field $E^{trial}(t)=\sin t$ as an initial guess 
at the first iteration step. The final optimized field 
$E(t)$ does not strongly depend on the trial field  for long-time control.
In this section, we investigate the influence of the initial trial field $E^{trial}(t)$ on 
the optimal field $E(t)$ for the short-time control as in Fig.8.
For that purpose,  as the initial state we use the delocalized state in the polychromatically
perturbed double-well potential ($M=10, \epsilon=1.0$). 
All numerical settings are same to the last section except for the first guess of the iteration.

\begin{figure}[h!]
\centering
\includegraphics[clip,scale=0.67]{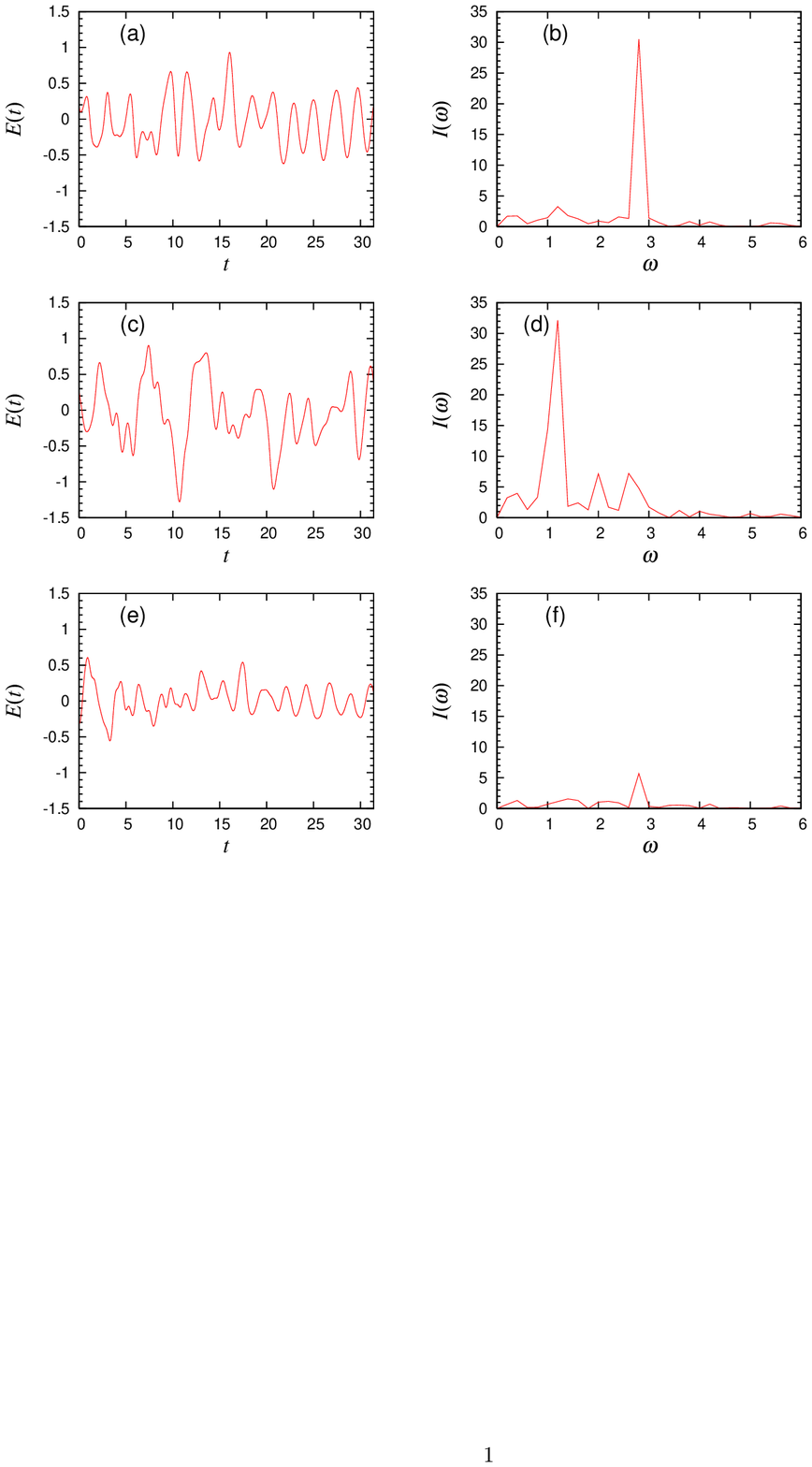}
\caption{
The trial field $E^{trial}(t)$ dependence of the short-time control  
between  the delocalized state  and the Gaussian wavepacket
in the polychromatically perturbed case ($M=10,\epsilon=1.0$). 
The left panels (a,c,e) are the optimized fields after 200 iterations 
for three different trial fields as the first guess, $E^{trial}_{2}(t)$, $E^{trial}_{3}(t)$ and $E^{trial}_{4}(t)$ 
given in the text. 
The right panels (b,d,f)  are the power spectra $I(\omega)$ corresponding to the optimal fields in the left panels.
We used $\alpha=0.1$  and $t_f=30$ in the calculation. 
As an initial state we used the initial delocalized state 
obtained (ug by) time-evolution up to $t=5.54 \times 10^3$ in the system with $M=10,\epsilon=1.0$.
The final overlaps are less than 0.8 in the all cases.
}
\label{fig11} 
\end{figure} 

\begin{figure}[h!]
\centering
\includegraphics[clip,scale=0.67]{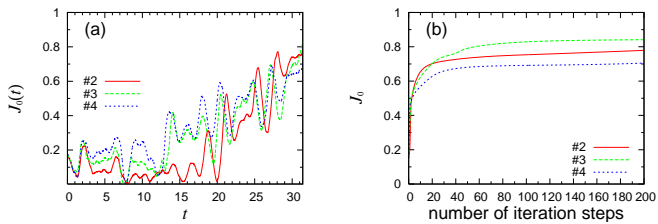}
\caption{
Comparison with the results for the different first guesses $E^{trial}(t)$ in Fig.\~11.
(a)Time-dependence of the squared final overlap $J_0(t)$ after 200 iterations.  
(b)Convergence of the optimized objective functional values  $J$ and $J_0$ as a function of the
number of iteration.
The other parameters are  the  same as a case in Fig.8.
The curves denoted by the figures ($2,3,4$) express the results corresponding to the
$E^{trial}_{2}(t)$, $E^{trial}_{3}(t)$ and $E^{trial}_{4}(t)$. 
}
\label{fig12} 
\end{figure} 

In Fig.11 we show several optimal fields for the short-time control in the transition 
between the delocalized state and the Gaussian wavepacket, where we used the following
trial  fields as the first guess,  $E^{trial}_{2}(t)=\sin 2.8 t$, 
$E^{trial}_{3}(t)=\frac{1}{\surd M} \sum_{i=1}^M \sin(\Omega_i  t )$,
$E^{trial}_{4}(t)=f(t)$. 
The $f(t)$ is randomly and uniformly distributed in the range $[-1,1]$. 
 As shown in  the  right panels of Fig.11 the shape of the power spectra 
corresponding to the these optimal fields are quite different with respect to one another. 
The function form of the optimal field $E(t)$ strongly depends on the trial function form 
$E^{trial}(t)$ as the first guess.  
Figure 12 shows the convergence to the optimal fields in the cases of 
the different initial fields $E^{trial}(t)$.
The optimized values $J_0$ after 200 iterations are about 0.8 in all cases, 
and the control is incomplete.

As a result, it is conjectured that during the optimization process the numerous local minima prevent the 
convergence to the global minimum corresponding to the target state.
This difficulty in finding  the global minimum comes from a feature of 
the chaos-induced delocalized state  with the {\it complicated phase structure} like an entangled state.
It is expected that 
the transition between the delocalized 
state and  the Gaussian wavepacket 
 can be achieved by the more complicated control due to existence of  a lot of local minima in the solution space
via the excited states with high energy than that between the Gaussian wavepackets. 
In the short-time control, for the sake of the manipulation of  the control field for the delocalized states, 
we need  some strategies in order to throw off the trap in the local minimum, 
such as genetic algorithm  \cite{sugawara01} and 
chaotic itinerancy \cite{kaneko90}.
Further investigation is necessary to find the global optimal field for short-time control problem 
concerning the chaos-induced delocalized state \cite{igarashi06c}.
The development of new optimal control method is out of scope in this paper.

\section{chaos-assisted control}
In this section, we consider influence of chaos in the controlled system $H_0(t)$ on the controllability 
for the delocalized state. 
All numerical settings are same to the last section 
except for the first guess of the iteration 
except for value of $\epsilon$ in $A(t)$.

As shown in Sect.II the chaotic behavior in the perturbed double-well 
system is enhanced by the increase of the perturbation strength $\epsilon$.

\begin{figure}[h!]
\centering
\includegraphics[clip,scale=0.67]{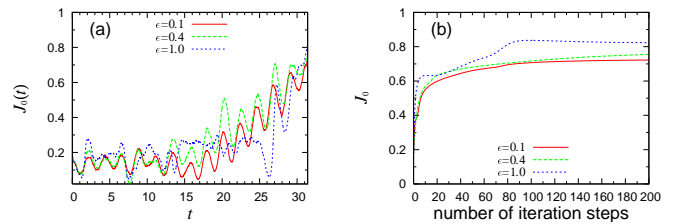}
\caption{
$\epsilon-$dependence of the short-time control in  the polychromatically perturbed case ($M=10$). 
(a)Time-dependence of the squared final overlap $J_0(t)$ after 200 iterations.  
(b)Convergence of the optimized objective functional value $J$ and $J_0$ as a function of the
number of iteration.
We used $\alpha=0.1$  and $t_f=30$ in the calculation. 
As an initial state we used the  delocalized state 
obtained  time-evolution up to $t=5.54 \times 10^3$ in the system with $M=10,\epsilon=1.0$.
}
\label{fig13} 
\end{figure} 

Figure 13 shows the short-time control for various values of  $\epsilon(=0.1, 0.4, 1.0)$ 
in  the polychromatically perturbed case ($M=10$). 
It found that the controllability is suppressed and the final overlap less converges to 
the target state as the decrease of the perturbation strength. 
We can regard that the strong chaotic mixing in the double-well system assists the control of the
transition between the quantum states. 
We call the phenomenon {\it chaos-assisted control}.

A simple scenario for chaos-assisted control  is attributed to the following fact.
In  a  weak chaotic system with small perturbation strength, the existence of the dynamical structure in 
the phase space can work as the obstacles for the motion of the wavepacket.
On the other hand, the strong perturbation can remove the restriction of the motion 
and make displacement of the wavepacket smoother than the weakly perturbed case.
The chaos-assisted control can be also observed in short-time control between Gaussian wavepackets
discussed in the last section. 

What  cases are uncontrollable even for long-target-time $t_f$?
The energy difference between the initial and target states may make difficult to 
control the transition of the delocalized state if the initial state is at least 
a nonstationary state with high energy.
We have confirmed the difficulty of the control even for a very long-target time 
optimization when we use a delocalized state with the higher energy as 
an initial state.
Then, it can be expected that the harder the control becomes
 the more efficient the chaos-assisted control may  work, 
although the control is still far from the perfect one. 
The details will be reported elsewhere in focusing on the more systematic 
investigation for this point \cite{igarashi06c}.

\section{Summary and Discussion}
   In summary,  we numerically investigated 
controllability of the
wave packet dynamics in the polychromatically perturbed 
one-dimensional double-well system based on  the optimal control theory.
The controllability depends on the target time $t_f$ and the initial state.
The results are summarized as follows.

(1) In the cases of the relatively long-time control for the displacement 
between the localized Gaussian wavepackets in the  wells,  
it can be well-controllable regardless of the perturbation type.

(2) In the cases of the relatively short-time control for the Gaussian wavepackets, 
we can obtain the almost  optimal field although the convergence of the iteration is slowly.

(3) When we use a delocalized state as the initial state, which is obtained by 
the time-evolution under the polychromatic perturbation ($M=10,\epsilon=1.0$), 
the quantum control becomes more insufficient than the case of the Gaussian wavepacket. 
The convergence of the iteration is slow because of numerous local minima.

(4) As the increase of the perturbation strength $\epsilon$ in the controlled system 
the controllability of the delocalized state can be enhanced because the restriction caused by the dynamical 
structure could be broken in the fully chaotic state.. 
We called the phenomenon {\it chaos-assisted control}.

It is interesting that whether or not the short-time optimization of 
the delocalized state is well-controllable by the other 
 control methods, for example, noniterative method and so on,  proposed
by some groups \cite{takami05,dannis03}. 

\vspace{1cm}

We modeled the other degrees of freedom coupled with the double-well system
as external coherent perturbation in Eq.\~(1). 
 A control  problem for a 
quantum system interacting with a bath composed of finite (or infinite) degrees of freedom
is interesting 
\cite{schrimer01,alessandro02}.
The application of OCT to laser driven hydrogen tunneling 
by using ultrafast laser pulses have been proposed by some groups 
\cite{naundorf99}. 
Then decoherence through coupling to intermolecular and intramolecular 
degrees of freedom is inevitable in the realistic system. 
The influence of the other degrees of freedom is modeled as coupling with 
an infinite collection of harmonic oscillators characterized by a broad and 
continuous distribution of frequencies.  
From the point of view, we modeled the effect of coupling with the other degrees of
freedom as a coupling with finite collection of highly excited harmonic oscillators
with some discrete special frequencies.
The control of the dissipative quantum state and entangled state  
are directly related to the realization of the quantum computation 
\cite{feynman96,kawabata03}.

The transfer mechanism of the well-localized Gaussian wavepacket 
between the wells has  an  analogy with  
Mott's variable range hopping (VRH) between the localized states in (ig
a) disordered system 
containing the roughness in the potential energy landscape,  
which is a dominant transport mechanism in low-temperature limit \cite{mott71}. 
On the other hand, the conductive properties of DNA have recently attracted interest
\cite{endres04,roche04,daphne05,gutierrez05}.
The lambda phase DNA ($\lambda-$DNA)
 has been sometimes modeled as a one-dimensional disordered system, and 
there has been an explanation for the electrical conductivity along the DNA double helix,  
by using the VRH based on the temperature dependence of the conductivity \cite{yu01}. 
 Therefore, the control of the electron transfer in the disordered system 
composed of fluctuating multi-well potential is also interesting problem.

 Moreover, we can realize the  another  delocalized state in the polychromatically perturbed 
one-dimensional disordered system \cite{yamada99}. 
    The more details of the controllability of the delocalized states in the disordered system and 
the double-well system 
  will be reported in the future \cite{igarashi06c}.

The authors would like to thank Prof. K.S. Ikeda and Prof. M. Goda for encouragements
and interests in this study.
We also would like to thank Dr. T. Takami for useful discussion in some meetings.  




\begin{thebibliography}{00}
\bibitem{frequency} The details of the correspondence has been given 
in \cite{yamada99}.
\end{thebibliography}
\end{document}